\documentclass[aps,pre,preprint, showpacs, superscriptaddress, showkeys]{revtex4}

\usepackage[dvips]{graphicx}
\usepackage{latexsym}
\usepackage{natbib}

\begin{document}


  \title{Transport in weighted networks: \\ Partition into superhighways and
    roads}

\author{Zhenhua Wu}
\affiliation{Center for Polymer Studies, Boston University, Boston,
  Massachusetts 02215, USA}
\author{Lidia A. Braunstein}
\affiliation{Departamento de F\'{\i}sica, Facultad de Ciencias Exactas y
  Naturales, Universidad Nacional de Mar del Plata, Funes 3350, 7600 Mar del
  Plata, Argentina}
\affiliation{Center for Polymer Studies, Boston University, Boston,
  Massachusetts 02215, USA}
\author{Shlomo Havlin}
\affiliation{Minerva Center of Department of Physics, Bar-Ilan
  University, Ramat Gan, Israel}
\author{H. Eugene Stanley}
\affiliation{Center for Polymer Studies, Boston University, Boston,
  Massachusetts 02215, USA}

\begin{abstract}

Transport in weighted networks is dominated by the minimum spanning tree
(MST), the tree connecting all nodes with the minimum total weight. We find
that the MST can be partitioned into two distinct components, having
significantly different transport properties, characterized by centrality ---
number of times a node (or link) is used by transport paths. One component,
the {\it superhighways}, is the infinite incipient percolation cluster; for
which we find that nodes (or links) with high centrality dominate. For the
other component, {\it roads}, which includes the remaining nodes, low
centrality nodes dominate. We find also that the distribution of the
centrality for the infinite incipient percolation cluster satisfies a power
law, with an exponent smaller than that for the entire MST. The significance
of this finding by showing that one can improve significantly the global
transport by improving a very small fraction of the network, the
superhighways.

\end{abstract}


\maketitle

Recently much attention has been focused on the topic of complex networks,
which characterize many natural and man-made systems, such as the internet,
airline transport system, power grid infrastructures, and the world wide
web~\cite{Barabasi_rmp_review, vespignani_book, Mendes_book}.
Besides the static properties of complex networks, dynamical phenomena such
as transport in networks are of vital importance from both theoretical and
practical perspectives. Recently much effort has been focused on weighted
networks~\cite{Barrat_WAN, Macdonald_ecoli}, where each link or node is
associated with a weight. Weighted networks yield a more realistic
description of real networks. For example, the cable links between computers
in the internet network have different weights, representing their capacities
or bandwidths.

In weighted networks the minimum spanning tree (MST) is a tree including all
of the nodes but only a subset of the links, which has the minimum total
weight out of all possible trees that span the entire network. Also, the MST
is the union of all ``strong disorder'' optimal paths between any two
nodes~\cite{FN_sd, Barabasi_ibp, Dobrin_mst, Cieplak:op1, Porto,
Lidia_op_prl, Zhenhua}. The MST which plays a major role for transport is
widely used in different fields, such as the design and operation of
communication networks, the traveling salesman problem, the protein
interaction problem, optimal traffic flow, and economic
networks~\cite{Khan_tech, Skiena_book_mst, Fredman_mst, Kruskal_mst,
Macdonald_ecoli, mst_eco_1, mst_eco_2}.

An important quantity that characterizes transport in networks is the
betweenness centrality, $C$, which is the number of times a node (or link)
used by the set of all shortest paths between all pairs of
nodes~\cite{Newman_centrality, Goh_load_prl, Kim_bc}. For simplicity we call
the ``betweenness centrality'' here ``centrality'' and we use the notation
``nodes'' but similar results have been obtained for links. The centrality,
$C$, quantifies the ``importance'' of a node for transport in the
network. Moreover, identifying the nodes with high $C$ enables, as shown
below, to improve their transport capacity and thus improve the global
transport in the network. The probability density function (pdf) of $C$ was
studied on the MST for both scale-free (SF)~\cite{Barabasi_sf} and
Erd\H{o}s-R\'{e}nyi (ER)~\cite{erdos} networks and found to satisfy a power
law,
\begin{equation}
  {\cal P}_{\rm MST}(C) \sim C^{-\delta_{\rm MST}}
\end{equation}
with $\delta_{\rm MST}$ close to $2$~\cite{Goh_centrality, Kim_bc}.

Here we show that a sub-network of the MST~\cite{FN_IIC_isIn_MST}, the
infinite incipient percolation cluster (IIC) has a significantly higher
average $C$ than the entire MST --- i.e., the set of nodes inside the IIC are
typically used by transport paths more often than other nodes in the MST. ---
In this sense the IIC can be viewed as a set of {\it superhighways} (SHW) in
the MST. The nodes on the MST which are not in the IIC are called {\it
roads}, due to their analogy with roads which are not superhighways (usually
used by local residents). We demonstrate the impact of this finding by
showing that improving the capacity of the superhighways (IIC) is
surprisingly a better strategy to enhance global transport compared to
improving the same number of links of the highest $C$ in the MST, although
they have higher $C$~\cite{FN_highest_bc_IIC}. This counterintuitive result
shows the advantage of identifying the IIC subsystem, which is very small
compared to the full network~\cite{FN_iic_mst_mass_ratio}. Our results are
based on extensive numerical studies for centrality of the IIC, and
comparison with the centrality of the entire MST. We study ER, SF and square
lattice networks.

To generate a ER network of size $N$ with average degree $\langle k \rangle$,
we pick at random a pair of nodes from all possible $N (N-1) /2$ pairs, link
this pair, and continue this process until we have exactly $\langle k \rangle
N/2$ edges. We disallow multiple connections between two nodes and self-loops
in a single node. To construct SF networks with a prescribed power law
distribution ${\cal P}(k) \sim k^{-\lambda}$ with $k \ge k_{\rm
min}$~\cite{Barabasi_sf}, we use the Molloy-Reed algorithm~\cite{Molloy_Reed,
Molloy_Reed_book}. We assign to each node $i$ a random number $k_i$ of links
drawn from this power law distribution. Then we choose a node $i$ and connect
each of its $k_i$ links with randomly selected $k_i$ different nodes.

To construct a {\it weighted} network, we next assign a weight $w_i$ to each
link from a uniform distribution between $0$ and $1$. The MST is obtained
from the weighted network using Prim's algorithm~\cite{network_flow_book}. We
start from any node in the largest connected component of the network and
grow a tree-like cluster to the nearest neighbor with the minimum weight
until the MST includes all the nodes of the largest connected component. Once
the MST is built, we compute the value of $C$ of each node by counting the
number of paths between all possible pairs passing through that node . We
normalize $C$ by the total number of pairs in the MST, $N(N-1)/2$, which
ensures that $C$ is between 0 and 1~\cite{FN_sd_mst}.

To find the IIC of ER and SF networks, we start with the fully connected
network and remove links in descending order of their weights. After each
removal of a link, we calculate $\kappa \equiv \langle k^2 \rangle /\langle k
\rangle$, which decreases with link removals. When $\kappa < 2 $, we stop the
process because at this point, the largest remaining component is the
IIC~\cite{Cohen_random_attack}. For the two dimensional (2D) square lattice
we cut the links in descending order of their weights until we reach the
percolation threshold, $p_c$ ($=0.5$). At that point the largest remaining
component is the IIC~\cite{Bunde_book}.

\begin{figure}
  \includegraphics[width=\textwidth]{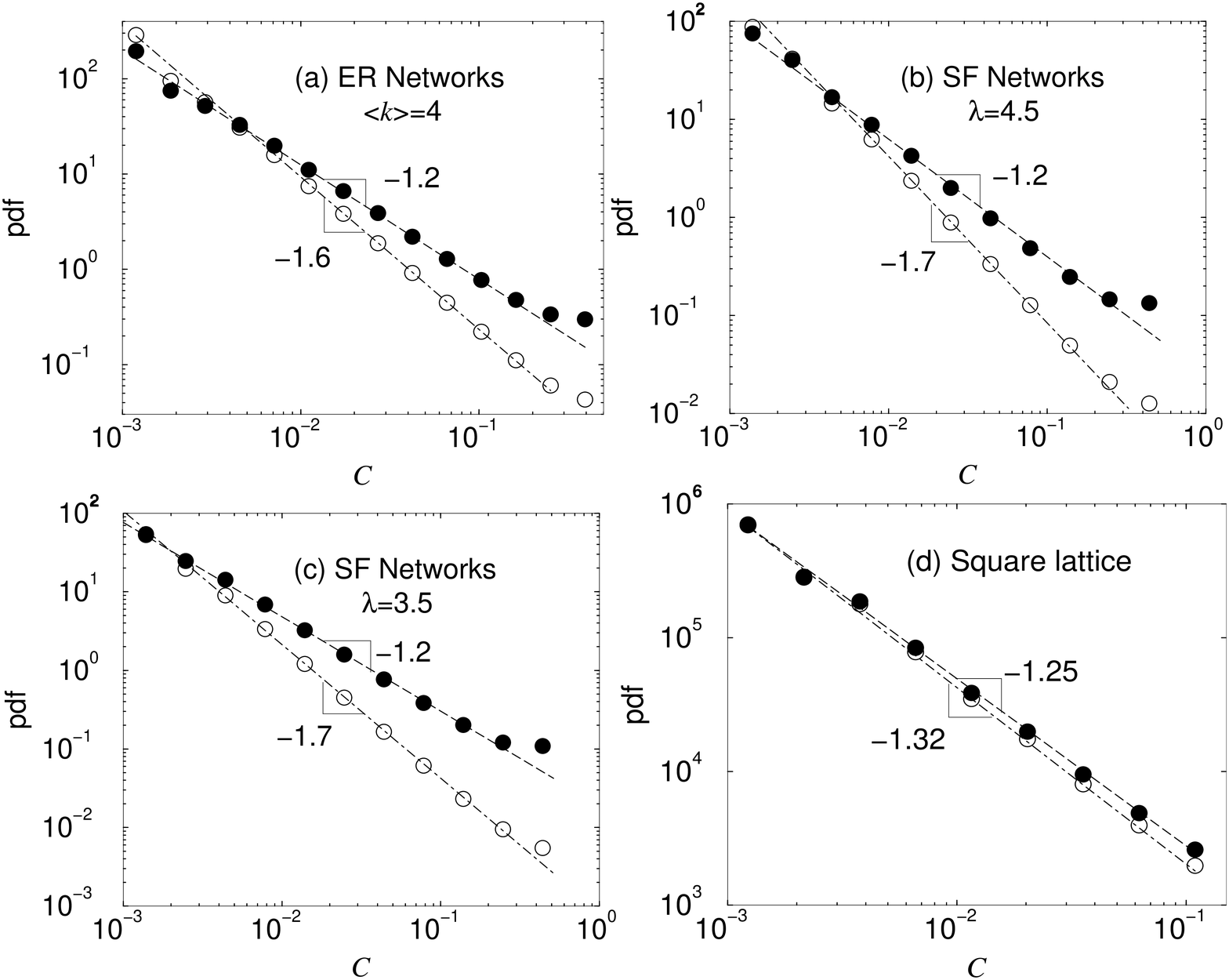}
  \caption{The pdf of the centrality of nodes for (a) ER graph with $\langle
    k \rangle = 4$, (b) SF with $\lambda = 4.5$, (c) SF with $\lambda = 3.5$
    and (d) $90 \times 90$ square lattice. For ER and SF, $N = 8192$ and for
    the square lattice $N = 8100$ . We analyze $10^4$ realizations. For each
    graph, the filled circles show ${\cal P}_{\rm IIC}(C)$; the unfilled
    circles show ${\cal P}_{\rm MST}(C)$.}
  \label{graph_Pcent_node}
\end{figure}

To quantitatively study the centrality of the nodes in the IIC, we calculate
the pdf, ${\cal P}_{\rm IIC}(C)$ of $C$. In Fig.~\ref{graph_Pcent_node} we
show for nodes that for all three cases studied, ER, SF and square lattice
networks, ${\cal P}_{\rm IIC}(C)$ satisfies a power law
\begin{equation}
  {\cal P}_{\rm IIC}(C) \sim C^{-\delta_{\rm IIC}},
  \label{BC_IIC_scaling}
\end{equation}
where 
\begin{equation}
  \delta_{\rm IIC} \approx \left\{ \begin{array}{ll}
    1.2  & {\rm [ER, SF] } \\
    1.25 & {\rm [square~lattice]}
  \end{array}\right..
\end{equation}
Moreover, from Fig.~\ref{graph_Pcent_node}, we find that $\delta_{\rm IIC} <
\delta_{\rm MST}$, implying a larger probability to find a larger value of
$C$ in the IIC compared to the entire MST. Our values for $\delta_{\rm MST}$
are consistent with those found in Ref.~\cite{Goh_centrality}. We obtain
similar results for the centrality of the links. Our results thus show that
the IIC is like a network of {\it superhighways} inside the MST. When we
analyze centrality for the entire MST, the effect of the high $C$ of the IIC
is not seen since the IIC is only a small fraction of the MST. Our results
are summarized in Table~\ref{table_para}.

%
%

\begin{figure}
  \includegraphics[width=\textwidth]{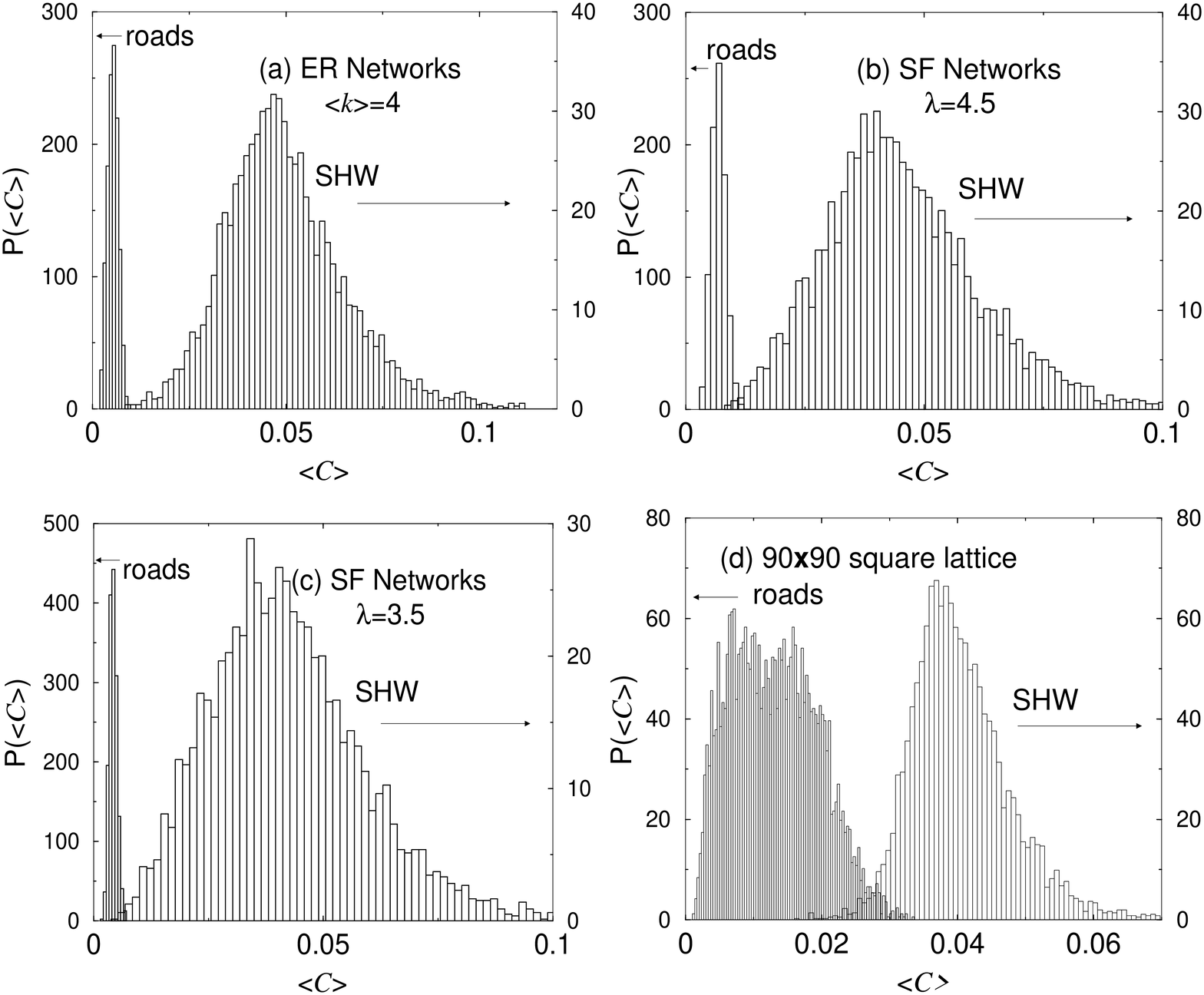}
  \caption{The normalized pdf for superhighway and roads of $\langle C
    \rangle$, the $C$ averaged over all nodes in one realization. (a) ER
    network, (b) SF network with $\lambda = 4.5$, (c) SF network with
    $\lambda = 3.5$ and (d) square lattice network. To make each histogram,
    we analyze 1000 network configurations.}
  \label{graph_ave_bc_node}
\end{figure}

To further demonstrate the significance of the IIC, we compute for each
realization of the network the average $C$ over all nodes, $\langle C
\rangle$. In Fig.~\ref{graph_ave_bc_node}, we show the histograms of $\langle
C \rangle$ for both the IIC and for the other nodes on the MST. We see that
the nodes on the IIC have a much larger $\langle C \rangle$ than the other
nodes of the MST.

%

\begin{figure}
  \includegraphics[width=0.6\textwidth]{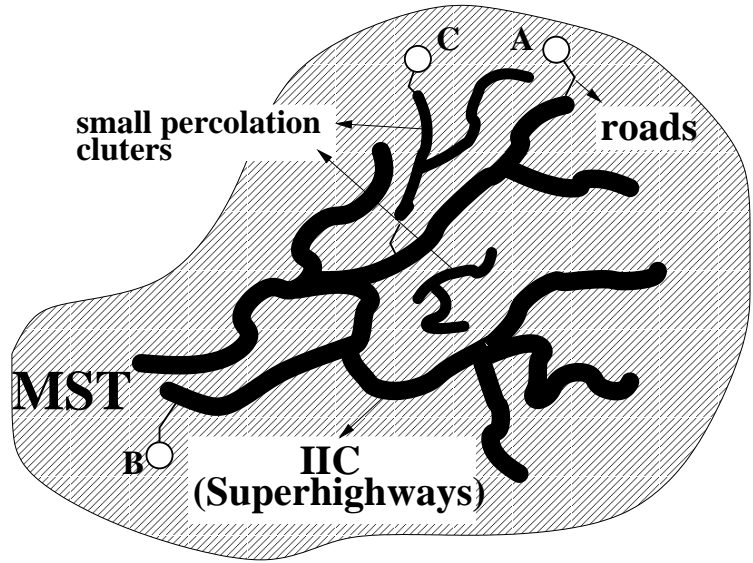}
  \caption{Schematic graph of the network of connected superhighways (heavy
    lines) inside the MST (shaded). A, B and C are examples of possible entry
    and exit nodes, which connect to the network of superhighways by
    ``roads'' (thin lines). The middle size lines indicates other percolation
    clusters with much smaller size compared to the IIC.}
  \label{graph_schematic_of_hw}
\end{figure}

\begin{figure}
  \includegraphics[width=\textwidth]{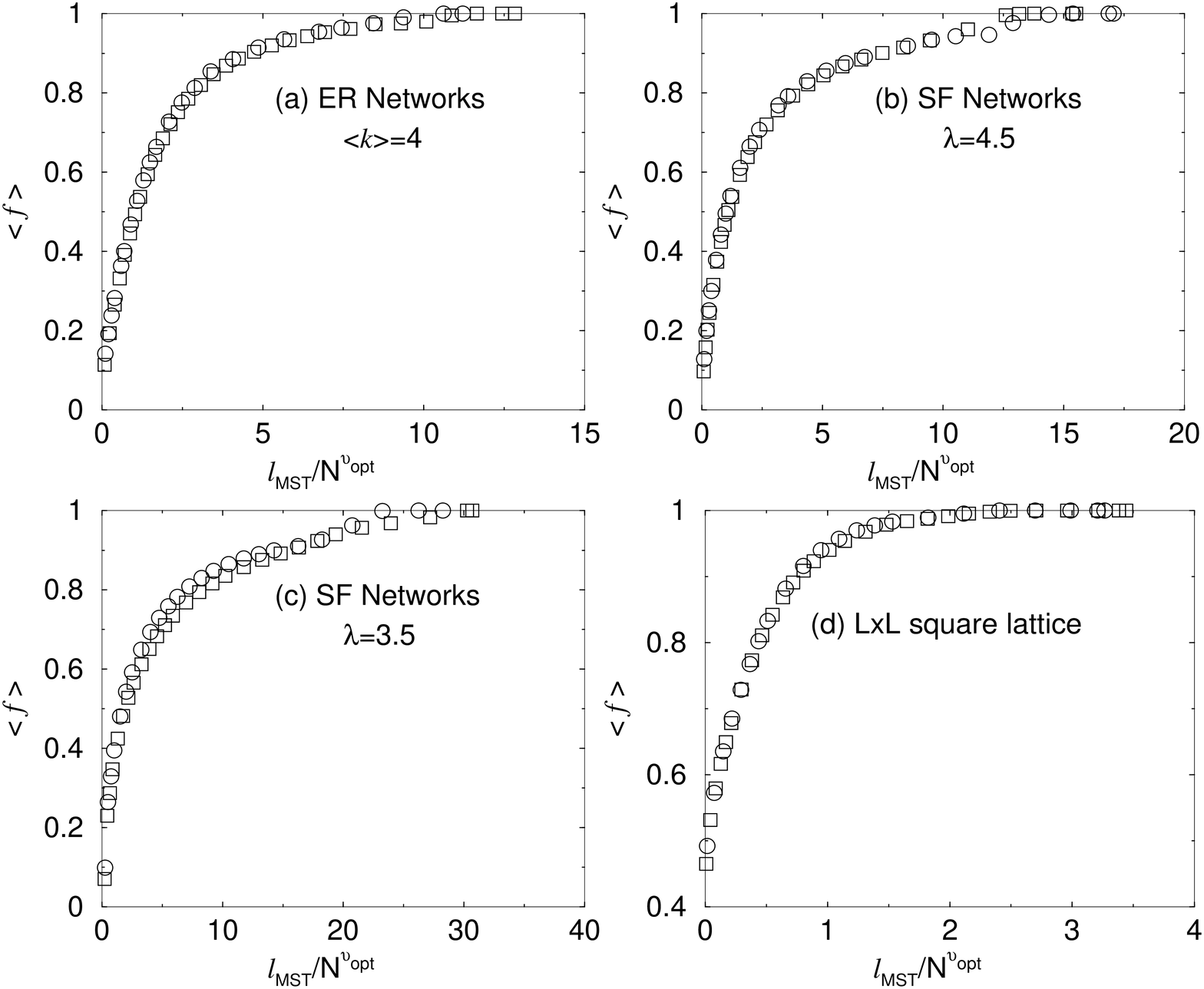}
  \caption{The average fraction, $\langle f \rangle$, of pairs using the SHW,
    as a function of $\ell_{\rm MST}$, the distance on the {\rm MST}. (a) ER
    graph with $\langle k \rangle = 4$, (b) SF with $\lambda = 4.5$, (c) SF
    with $\lambda = 3.5$ and (d) square lattice. For ER and SF:
    ($\bigcirc$)$N = 1024$ and ($\Box$)$N = 2048$ with $10^4$
    realizations. For square lattice: ($\bigcirc$)$N = 1024$ and ($\Box$)$N =
    2500$ with $10^3$ realizations. The $x$ axis is rescaled by $N^{\nu_{\rm
    opt}}$, where $\nu_{\rm opt} = 1/3$ for ER and for SF with $\lambda > 4$,
    and $\nu_{\rm opt} = (\lambda - 3)/(\lambda -1)$ for SF networks with $3
    < \lambda < 4$~\cite{Lidia_op_prl}. For the $L \times L$ square lattice,
    $\ell_{\rm MST} \sim L^{d_{\rm opt}}$ and since $L^2 = N$, $\nu_{\rm opt}
    = d_{\rm opt} / 2 \approx 0.61$~\cite{Cieplak:op1, Porto}.}
  \label{graph_ave_frac}
\end{figure}

Figure~\ref{graph_schematic_of_hw} shows a schematic plot of the SHW inside
the MST and demonstrates its use by the path between pairs of nodes. The MST
is the ``skeleton'' inside the network, which plays a key role in transport
between the nodes. However, the IIC in the MST is like the ``spine in the
skeleton'', which plays the role of the superhighways inside a road
transportation system. A car can drive from the entry node ${\rm A}$ on roads
until it reaches a superhighway, and finds the exit which is closest to the
exit node ${\rm B}$. Thus those nodes which are far from each other in the
MST should use the IIC superhighways more than those nodes which are close to
each other. In order to demonstrate this, we compute $f$, the average
fraction of pairs of nodes using the {\rm IIC}, as a function of $\ell_{\rm
MST}$, the distance between a pair of nodes on the MST
(Fig.~\ref{graph_ave_frac}). We see that $f$ increases and approaches one as
$\ell_{\rm MST}$ grows. We also show that $f$ scales as $\ell_{\rm MST} /
N^{\nu_{\rm opt}}$ for different system sizes, where $\nu_{\rm opt}$ is the
percolation connectedness exponent~\cite{Lidia_op_prl, Zhenhua}.



\begin{figure}
  \includegraphics[width=\textwidth]{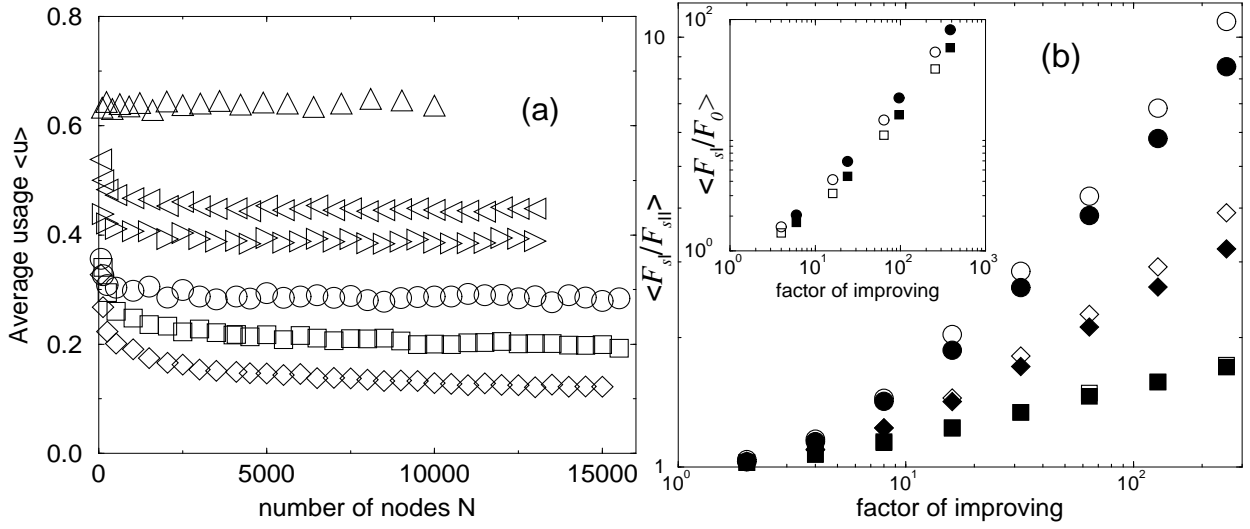}
  \caption{(a) The average usage $\langle u \rangle \equiv \langle \ell_{\rm
    IIC} / \ell_{\rm MST} \rangle$ for different networks, as a function of
    the number of nodes $N$. $\bigcirc$ (ER with $\langle k \rangle = 4$),
    $\Box$ (SF with $\lambda = 4.5$), $\Diamond$ (SF with $\lambda = 3.5$),
    $\bigtriangleup$ ($L \times L$ square lattice). The symbols ($\rhd$) and
    ($\lhd$) represent the average usage for ER with $\langle k \rangle = 4$
    when the two largest percolation clusters and the three largest
    percolation clusters are taken into account, respectively. (b) The ratio
    between the flow using strategy I, $F_{s\rm I}$,
    and that using strategy II, $F_{s\rm II}$, as a function of the
    factor of improving conductivity/capacity. The inset is 
    the ratio between the flow using strategy I and the flow in the original
    network, $F_{\rm 0}$. The data are all for ER networks
    with $N = 2048$, $\langle k \rangle = 4$ and $n = 50$($\bigcirc$), $n =
    250$($\Diamond$) and $n = 500$($\Box$). The unfilled symbols are for
    current flow and the filled symbols are for maximum flow.}
  \label{graph_ratio_and_flow}
\end{figure}

The next question is how much the IIC is used in transport on the MST? We
define the IIC {\it superhighway usage},
\begin{equation}
  u \equiv \frac{\ell_{\rm IIC}}{\ell_{\rm MST}},
\end{equation}
where $\ell_{\rm IIC}$ is the number of the links in a given path of length
$\ell_{\rm MST}$ belonging to the {\rm IIC} superhighways. The average usage
$\langle u \rangle$ quantifies how much the IIC is used by the transport
between all pairs of nodes. In Fig.~\ref{graph_ratio_and_flow}(a), we show
$\langle u \rangle$ as a function of the system size $N$. Our results suggest
that $\langle u \rangle$ approaches a constant value and becomes independent
of $N$ for large $N$. This is surprising since the average value of the ratio
between the number of nodes on the IIC and on the MST, $\langle N_{\rm IIC} /
N_{\rm MST} \rangle$, approaches zero as $N \to
\infty$~\cite{FN_iic_mst_mass_ratio}, showing that although the IIC contains
only a tiny fraction of the nodes in the entire network, its usage for the
transport in the entire network is constant. We find that $\langle u \rangle
\approx 0.3$ for ER networks, $\langle u \rangle \approx 0.2$ for SF networks
with $\lambda = 4.5$, and $\langle u \rangle \approx 0.64$ for the square
lattice. The reason why $\langle u \rangle$ is not close to $1.0$ is that in
addition to the IIC, the optimal path passes through other percolation
clusters, such as the second largest and the third largest percolation
clusters. In Fig.~\ref{graph_ratio_and_flow}, we also show for ER networks,
the average usage of the two largest and the three largest percolation
clusters for a path on the MST and we see that the average usage increases
significantly and is also independent of $N$. However, the number of clusters
used by a path on MST is relatively small and proportional to $\ln
N$~\cite{Sameet_mst}, suggesting that the path on the MST uses only a few
percolation clusters and a few jumps between them ($\sim \ln N$) in order to
get from an entry node to an exit node on the network. When $N \to \infty$
the average usage of all percolation clusters should approach $1$.


Can we use the above results to improve the transport in networks? It is
clear that by improving the capacity or conductivity of the highest $C$ links
one can improve the transport (see Fig.~\ref{graph_ratio_and_flow}(b)
inset). We hypothesize that improving the IIC links (strategy I), which
represent the superhighways is more effective than improving the same number
of links with the highest $C$ in the MST (strategy II), although they have
higher centrality~\cite{FN_highest_bc_IIC}. To test the hypothesis, we study
two transport problems: (i) current flow in random resistor networks, where
each link of the network represents a resistor and (ii) the maximum flow
problem well known in computer science~\cite{Algorithm_book}. We assign to
each link of the network a resistance/capacity, $e^{ax}$, where $x$ is an
uniform random number between 0 and 1, with $a = 40$. The value of $a$ is
chosen such as to have a broad distribution of disorder so that the MST
carries most of the flow~\cite{Zhenhua, Sameet_mst}. We randomly choose $n$
pairs of nodes as sources and other $n$ nodes as sinks and compute flow
between them. We compare the transport by improving the conductance/capacity
of the links on the IIC (strategy I) with that by improving the same number
of links but those with the highest $C$ in the MST (strategy II). Since the
two sets are not the same and therefore higher centrality links will be
improved in II~\cite{FN_highest_bc_IIC}, it is tempting to suggest that the
better strategy to improve global flow would be strategy II. However, here we
demonstrate using ER networks as an example that counterintuitively strategy
I is better. We also find similar improvements of strategy I compared to
strategy II for SF networks with $\lambda = 3.5$. In
Fig.~\ref{graph_ratio_and_flow}(b), we compute the ratio between the flow
using strategy I ($F_{s\rm I}$) and the flow using strategy II ($F_{s\rm
II}$) as a function of the factor of improving conductivity/capacity of the
links. The figure clearly shows that strategy I is better than strategy
II. Since the number of links in the IIC is relatively very small comparing
to the number of links in the whole network~\cite{FN_iic_mst_mass_ratio}, it
could be a very efficient strategy.

In summary, we find that the centrality of the IIC for transport in networks
is significantly larger than the centrality of the other nodes in the
MST. Thus the IIC is a key component for transport in the MST. We demonstrate
that improving the capacity/conductance of the links in the IIC is useful
strategy to improve transport.

We thank ONR, Israel Science Foundation, European NEST project DYSONET,
FONCyT (PICT-O2004/370) and Israeli Center for Complexity Science for
financial support.


\begin{thebibliography}{99}
\bibitem{FN_sd} By strong disorder it is meant that the weight of the path
    is determined by the largest weight along the path.
\bibitem{FN_IIC_isIn_MST} The IIC contains loops in lattices in dimension
    $d$ below 6. However, for networks ($d = \infty$), the IIC contains no
    loops and in this case for large $N$, the IIC must be a subset of the
    MST. In our simulations, we tested this and found that more than $99\%$
    links of IIC belong to MST. For lattices, we only choose the part of the
    IIC that belongs to the MST.
\bibitem{FN_highest_bc_IIC} The overlap between the two groups is about
    $30\%$ for ER networks.
\bibitem{FN_sd_mst} This $C$ measurement is equivalent to counting the
    number of times a node (link) is used by the set of optimal paths linking
    all pairs of nodes, in the limit of strong disorder.
\end{thebibliography}

\newpage

\begin{table}[!h]
  \begin{tabular}{|c|c|c|c|c|}
    \hline                       & ER  & SF ($\lambda = 4.5$) & SF ($\lambda = 3.5$) & square lattice \\
    \hline $\delta_{\rm IIC}$    & 1.2 & 1.2 & 1.2 & 1.25 \\
    \hline $\delta_{\rm MST}$    & 1.6 & 1.7 & 1.7 & 1.32 \\
    \hline $\nu_{\rm opt}$       & 1/3 & 1/3 & 0.2 & 0.61 \\
    \hline $\langle u \rangle$   & $0.29$ & $0.20$ & $0.13$ & $0.64$ \\
    \hline
  \end{tabular}
  \caption{Results for the IIC and the MST}
  \label{table_para}
\end{table}

\newpage

\newpage

\newpage

\newpage

\newpage

\end{document}